\documentclass[12pt]{article}
\usepackage{amsfonts}
\usepackage{amssymb}
\usepackage{graphics,psboxit,amsmath} 
\usepackage{subfigure}
\usepackage{graphicx}
\usepackage{verbatim}

\def\hybrid{\topmargin -30pt    \oddsidemargin 0pt 
        \headheight 0pt \headsep 0pt
        \textwidth 6.25in       
        \textheight 9.5in       
        \marginparwidth .875in
        \parskip 5pt plus 1pt   \jot = 1.5ex}

\hybrid

\def\baselinestretch{1.2}

\catcode`\@=11

\def\marginnote#1{}
\newcount\hour
\newcount\minute
\newtoks\amorpm
\hour=\time\divide\hour by60
\minute=\time{\multiply\hour by60 \global\advance\minute by-\hour}
\edef\standardtime{{\ifnum\hour<12 \global\amorpm={am}%
        \else\global\amorpm={pm}\advance\hour by-12 \fi
        \ifnum\hour=0 \hour=12 \fi
        \number\hour:\ifnum\minute<10 0\fi\number\minute\the\amorpm}}
\edef\militarytime{\number\hour:\ifnum\minute<10 0\fi\number\minute}

\def\draftlabel#1{{\@bsphack\if@filesw {\let\thepage\relax
   \xdef\@gtempa{\write\@auxout{\string
      \newlabel{#1}{{\@currentlabel}{\thepage}}}}}\@gtempa
   \if@nobreak \ifvmode\nobreak\fi\fi\fi\@esphack}
        \gdef\@eqnlabel{#1}}
\def\@eqnlabel{}
\def\@vacuum{}
\def\draftmarginnote#1{\marginpar{\raggedright\scriptsize\tt#1}}

\def\draft{\oddsidemargin -.5truein
        \def\@oddfoot{\sl preliminary draft \hfil
        \rm\thepage\hfil\sl\today\quad\militarytime}
        \let\@evenfoot\@oddfoot \overfullrule 3pt
        \let\label=\draftlabel
        \let\marginnote=\draftmarginnote
   \def\@eqnnum{(\theequation)\rlap{\kern\marginparsep\tt\@eqnlabel}%
\global\let\@eqnlabel\@vacuum}  }

\def\draft2{
        \def\@oddfoot{\sl preliminary draft \hfil
        \rm\thepage\hfil\sl\today\quad\militarytime}
        \let\@evenfoot\@oddfoot \overfullrule 3pt
        \let\label=\draftlabel
        \let\marginnote=\draftmarginnote
   \def\@eqnnum{(\theequation)\rlap{\kern\marginparsep\tt\@eqnlabel}%
\global\let\@eqnlabel\@vacuum}  }

\def\preprint{\twocolumn\sloppy\flushbottom\parindent 2em
        \leftmargini 2em\leftmarginv .5em\leftmarginvi .5em
        \oddsidemargin -.5in    \evensidemargin -.5in
        \columnsep .4in \footheight 0pt
        \textwidth 10.in        \topmargin  -.4in
        \headheight 12pt \topskip .4in
        \textheight 6.9in \footskip 0pt
        \def\@oddhead{\thepage\hfil\addtocounter{page}{1}\thepage}
        \let\@evenhead\@oddhead \def\@oddfoot{} \def\@evenfoot{} }

\def\numberbysection{\@addtoreset{equation}{section}
        \def\theequation{\thesection.\arabic{equation}}}

\def\underline#1{\relax\ifmmode\@@underline#1\else
        $\@@underline{\hbox{#1}}$\relax\fi}

\def\titlepage{\@restonecolfalse\if@twocolumn\@restonecoltrue\onecolumn
     \else \newpage \fi \thispagestyle{empty}\c@page\z@
        \def\thefootnote{\fnsymbol{footnote}} }

\def\endtitlepage{\if@restonecol\twocolumn \else \newpage \fi
        \def\thefootnote{\arabic{footnote}}
        \setcounter{footnote}{0}}  

\catcode`@=12
\relax

\def\figcap{\section*{Figure Captions\markboth
        {FIGURECAPTIONS}{FIGURECAPTIONS}}\list
        {Figure \arabic{enumi}:\hfill}{\settowidth\labelwidth{Figure
999:}
        \leftmargin\labelwidth
        \advance\leftmargin\labelsep\usecounter{enumi}}}
 \relax
\def\tablecap{\section*{Table Captions\markboth
        {TABLECAPTIONS}{TABLECAPTIONS}}\list
        {Table \arabic{enumi}:\hfill}{\settowidth\labelwidth{Table
999:}
        \leftmargin\labelwidth
        \advance\leftmargin\labelsep\usecounter{enumi}}}
 \relax
\def\reflist{\section*{References\markboth
        {REFLIST}{REFLIST}}\list
        {[\arabic{enumi}]\hfill}{\settowidth\labelwidth{[999]}
        \leftmargin\labelwidth
        \advance\leftmargin\labelsep\usecounter{enumi}}}
 \relax
\makeatletter
\newcounter{pubctr}
\def\publist{\@ifnextchar[{\@publist}{\@@publist}}
\def\@publist[#1]{\list
        {[\arabic{pubctr}]\hfill}{\settowidth\labelwidth{[999]}
        \leftmargin\labelwidth
        \advance\leftmargin\labelsep
        \@nmbrlisttrue\def\@listctr{pubctr}
        \setcounter{pubctr}{#1}\addtocounter{pubctr}{-1}}}
\def\@@publist{\list
        {[\arabic{pubctr}]\hfill}{\settowidth\labelwidth{[999]}
        \leftmargin\labelwidth
        \advance\leftmargin\labelsep
        \@nmbrlisttrue\def\@listctr{pubctr}}}
 \relax
\makeatother

\def\be{\begin{equation}}
\def\ee{\end{equation}}
\def\ba{\begin{eqnarray}}
\def\ea{\end{eqnarray}}

\def\del{\partial}

\def\k{\kappa}
\def\r{\rho}
\def\a{\alpha}

\def\b{\beta}

\def\G{\Gamma}
\def\d{\delta}
\def\D{\Delta}
\def\e{\epsilon}

\def\Th{\Theta}
\def\m{\mu}
\def\n{\nu}
\def\om{\omega}
\def\Om{\Omega}
\def\l{\lambda}
\def\L{\Lambda}
\def\s{\sigma}

\def\cL{{\cal L}}

\def\no{\noindent}

\def\qq{\qquad}

\def\IR{\relax{\rm I\kern-.18em R}}

\def\tL{{\tilde L}}
\def\PL{Poisson--Lie T-duality}

\def\inv{^{\raise.0ex\hbox{${\scriptscriptstyle -}$}\kern-.05em 1}}

\def \ha {{1\over 2}}

\def \ov {\over}

\begin{document}


\renewcommand{\theequation}{\thesection.\arabic{equation}}
\csname @addtoreset\endcsname{equation}{section}

\newcommand{\eqn}[1]{(\ref{#1})}
\begin{titlepage}
\begin{center}

{}\hfill QMUL-PH-09-22

\phantom{xx}
\vskip 0.5in

{\large \bf Renormalization of Lorentz non-invariant actions }\\
{\large \bf and manifest T-duality}

\vskip 0.5in

{\bf K. Sfetsos}${}^{1a}$,\phantom{x} {\bf K. Siampos}${}^{1b}$\phantom{x}
and \phantom{x} {\bf Daniel C. Thompson}${}^{2c}$
\vskip 0.1in

${}^1$Department of Engineering Sciences, University of Patras,\\
26110 Patras, Greece\\

\vskip .2in

${}^2$Queen Mary University of London, Department of Physics, \\
Mile End Road,
London, E1 4NS, United Kingdom\\

\end{center}

\vskip .4in

\centerline{\bf Synopsis}

\no
We study general two-dimensional $\s$-models which do not possess
manifest Lorentz invariance. We show how demanding that Lorentz
invariance is  recovered as an emergent on-shell symmetry constrains
these $\s$-models. The resulting actions have an
underlying group-theoretic structure and resemble
Poisson--Lie T-duality invariant actions. We consider the
one-loop renormalization of these  models and show that the quantum
Lorentz anomaly is absent. We calculate the running of the couplings
in general and show, with certain non-trivial examples, that this agrees with that of the T-dual
models obtained classically from the duality invariant action.
Hence, in these cases
solving constraints before and after quantization are commuting operations.

\vfill
\no
 {
 $^a$sfetsos@upatras.gr,\phantom{x}
 $^b$siampos@upatras.gr,\phantom{x} $^c$d.c.thompson@qmul.ac.uk}

\end{titlepage}
\vfill
\eject


\tableofcontents

\def\baselinestretch{1.2}
\baselineskip 20 pt
\no

\section{Introduction}

In theories possessing duality symmetries it is quite standard to lose manifest Lorentz
(and even reparameterization)  invariance
when they are formulated in a duality invariant way at the level of the action.
Common characteristic features of these theories
are the doubling of fields, the replacement of the Lorentz
by some other symmetry as well as the recovery of the usual Lorenz symmetry on-shell
(see \cite{FloJa,TseDual,SchSe,Duff}, for notable examples of this).
In addition, in such duality invariant actions it is possible to use the equations of motion to eliminate half of the fields resulting in a Lorentz invariant action.
If this can be done in more than one way, the corresponding theories are duality equivalent.
There are two issues that immediately arise which are actually related. Firstly, it is
important to know whether such an equivalence holds beyond the classical level.
Secondly, is the procedure of quantization
before eliminating extra fields equivalent to first eliminating fields and then quantizing?

\no
These issues can be addressed successfully with a generalization of the well known
abelian \cite{BUSCHER} and non-abelian \cite{nonabel} T-dualities, namely the so-called Poisson-Lie
T-duality \cite{KliSevI}.  Its most notable feature is that it does not
rely on the existence of isometries but rather on a rigid group-theoretical structure \cite{KliSevI}
known as the Drinfeld Double.  Nevertheless, it shares some common features with ordinary T-duality.
For instance, it can be explicitly formulated as a canonical transformation between phase-space
variables \cite{PLsfe1,PLsfe2}, similarly to ordinary T-duality
\cite{zacloz, AALcan}.

\no
Pairs of Poisson--Lie T-dual $\sigma$-models can be constructed classically from a single
duality invariant theory consisting of a WZW model on the Drinfeld Double supplemented with
a non-linear interaction term \cite{DriDou}. By choosing a parametrization for a group element
of the Double and implementing the constraints that arise as equations of motion,
one can eliminate half of the fields and obtain a standard Lorentz invariant $\sigma$-model.
The T-dual partner to this is obtained by choosing an inequivalent parametrization
for the group element and repeating the constraint procedure.
A first step in understanding whether Poisson--Lie T-duality holds beyond the classical
level was made in \cite{PLsfe3} where it was shown,
for specific examples, that the system of
renormalization group (RG) equations for couplings occurring in each of the T-dual theories
are equivalent at one-loop. The one-loop renormalization of general
Poisson--Lie T-dual models was considered in  \cite{ValCli} and a proof of
the equivalence of their RG equations was given in \cite{KSsquarePL}.

\no
It is interesting to ask whether these RG equations can be recast in
a manifestly duality invariant form.  For this to be so we should hope
that the duality invariant theory also produces the same system of RG equations.
This need not be the case since the process of constraining used to arrive at the
pair of T-dual models may not commute with the process of quantization.
This motivates our study of the one-loop renormalization  of the T-duality
invariant theory in particular and of Lorentz non-invariant $\s$-models in general.

\no
The organization of this paper is as follows:
In section 2 we consider general two-dimensional
$\sigma$-models that do not possess Lorentz invariance.  We show that
demanding Lorentz invariance as an emergent on-shell symmetry severely constrains
the background of the $\sigma$-model.  In fact, solving these constraints results
in an action that has the form of the Poisson--Lie T-duality action described above.

\no
In section 3 we consider the quantum behavior of these Lorentz
non-invariant theories  by calculating their one-loop effective actions using the
background field method.
The renormalization of similar theories was considered  \cite{Berman1, Berman2}
in the related context of the Doubled Formalism of abelian T-duality 
\cite{hull1} (a larger class of such doubled models has been considered at the classical level
in \cite{Dall'Agata}).
In our case however, we must extend this analysis to account for the non-trivial
group geometry of the group manifold.
One might expect both a Weyl anomaly and a Lorentz anomaly to occur.
We find that the counter term for the Weyl divergences can be absorbed
into a redefinition of the coupling constants and hence the models are
one-loop renormalizable. Furthermore we show that the potential Lorentz
anomaly vanishes through non-trivial cancellations.
This indicates quantum consistency of the Lorentz
non-invariant theories considered.

\no
In section 4 we apply our general construction to the Poisson--Lie T-duality invariant $\sigma$-model i.e.
 the case when the group manifold is a Drinfeld Double.
In this case the explicit expressions we construct for the
 Weyl anomaly provide us a manifestly duality invariant
description of the RG equations derived previously.
For specific non-trivial examples
we check that these do indeed reproduce the RG equations
obtained before \cite{PLsfe4,KSsquarePL}.

\no
We conclude our paper with section 5.

\section{On-shell Lorentz invariant actions}

In this section we develop the classical formalism for two-dimensional $\s$-models to possess on-shell
Lorentz invariance. For a related treatment the reader is also referred to \cite{Dall'Agata}.

\no
Consider a general Lorentz non-invariant two-dimensional bosonic $\s$-model action
\cite{TseDual}\footnote{In our conventions, the light-cone world-sheet
variables are $\s^\pm= \ha(\tau\pm \s)$ and hence $\del_\pm = \del_0 \pm
\del_1 $, where $\del_0 = \del_\tau$ and $\del_1= \del_\s$. The two-dimensional
world-sheet metric is diagonal with $\eta_{00}=-\eta_{11}=1$ and
also $\e^{01}=1$.}
\be
S= \ha \int d\s d\tau \left( C_{MN} \del_0 X^M \del_1 X^N + M_{MN} \del_1 X^M
\del_1 X^N \right) ~ ,
\label{act1}
\ee
where the general matrix $C_{MN}$ and the
symmetric matrix $M_{MN}$ depend on the $X^M$'s.
It is straightforward to show that the Lorentz transformations
\be
\d X^M = -\s \del_\tau X^M -\tau \del_\s X^M \ ,
\label{Lllor}
\ee
do not leave \eqn{act1} invariant, instead they result in
\be
\d_{\rm Lorentz} S = \int d\s d\tau \left( \ha S_{MN}(\del_1 X^M \del_1 X^N + \del_0 X^M \del_0 X^N)
+ M_{MN} \del_0 X^M \del_1 X^N \right)\ ,
\label{tralor}
\ee
where $S_{MN}$ is the symmetric part of $C_{MN}$, i.e. $S_{MN}=\ha (C_{MN}+
C_{NM})$.
We shall find it convenient to split the index $M$ such that $X^M=(X^\m,Y^i)$.  We would like to find the conditions under which \eqn{act1}
is on-shell Lorentz invariant,\footnote{Similar considerations have been made in \cite{TseDual} for constant
matrices $C$ and $M$.}
 using the equations of
motion corresponding to the $X^\m$'s. Since the equations of motion
for the $Y^i$'s will not be used, we may add to the
Lagrangian corresponding to \eqn{act1} any Lorentz invariant term $Q_{ij}
\del_+ Y^i \del_- Y^j$, for some matrix $Q_{ij}$ that may depend on the
$Y^i$'s. The fields $Y^i$ are called spectators.
The variation of \eqn{act1} with respect to these fields is
\be
\d S = \int d\s d\tau \ \d X^\m \left(-\del_1 E_\m + \ha \del_\m M_{N\L}
\del_1 X^N \del_1 X^\L + \hat \G_{N;\m\L} \del_0 X^N \del_1 X^\L \right) ~ ,
\label{varact1}
\ee
where
\be
E_\m = S_{\m N} \del_0 X^N + M_{\m N} \del_1 X^N \ ~ ,
\label{eqEm}
\ee
and
\be
\hat\G_{N;\m \L} = \ha (\del_\m C_{N\L} + \del_\L C_{\m N} -\del_N C_{\m\L})~ ,
\label{gmnl}
\ee
which as we shall soon see, when it is appropriately restricted, plays the r\^ole
of connection.

\no
Since \eqn{eqEm} and the equations of motions
are first and second order in world-sheet derivatives, respectively,
we should require that the latter can be written in the form $\del_1(\dots)$ in order to get
conditions that can be used to make \eqn{tralor} vanish.
However, the last two terms in \eqn{varact1} cannot be written immediately
in this form. Had this been the case performing the
derivative would result into second order derivatives like $\del^2_1 $ and $\del_0 \del_1 $
which do not appear in \eqn{varact1}.

\no
To proceed we require that the last two terms in \eqn{eqEm} can be cast into the form
\be
-\del_1 \L^\n_A \L^A_\m E_\n\ ,
\label{assup}
\ee
for some $X^\m$ dependent square matrix $\L^A_\m$ and its inverse $\L_A^\m$.
This is a very stringent condition with
severe consequences that restrict the type of backgrounds in the $\s$-models
that can be finally admitted by requiring
that Lorentz invariance emerges on-shell.
Then, assuming that \eqn{assup} holds, we find that \eqn{varact1} becomes
\be
\d S = -\int d\s d\tau  \ \d X^\m\left (\del_1 E_\m +\del_1 \L^\n_A \L^A_\m E_\n\right) =
-\int d\s d\tau  \ \d X^\m \L^A_\m \del_1 (\L^\n_A E_\n)\ .
\label{jsg2}
\ee
Hence, the equations of motion can be integrated once and read
\be
\L^\n_A E_\n = f_A(\tau)\ ,
\label{invars}
\ee
where the $f_A(\tau)$'s are otherwise arbitrary. However \eqn{jsg2} shows that
\eqn{act1} has a ``small'' local symmetry under the transformation
\be
\d X^\m = \L^\m_A \e^A(\tau)\ ,
\ee
where $\e^A(\tau)$ are some $\tau$-dependent parameters. This can be used
to set $f^A(\tau)=0$, $\forall\ A$, showing that the equations of motion are
first order and simply read
\be
E_\m = 0  ~ ,
\label{eqEm2}
\ee
where the $E_\m$'s are given by \eqn{eqEm}.

\subsection{Recovering on-shell Lorentz invariance}

Using \eqn{eqEm2}, we may solve for $\del_0 X^\m$ and after some algebraic manipulations recast the
anomalous Lorentz variation \eqn{tralor} into the form
\ba
&& \d_{\rm Lorentz} S\Big |_{\rm on \ shell}  =  \int d\s d\tau \Big[\ha (S_{MN} - M_{M\a}S^{\a\b} M_{\b N}) \del_1 X^M \del_1 X^N
\nonumber\\
&& + (M_{iN} -S_{i\a} S^{\a\b}M_{\b N})\del_0 Y^i \del_1 X^N + \ha (S_{ij} -S_{i\a} S^{\a\b} S_{\b j})
\del_0 Y^i \del_0 Y^j  \Big]\ ,
\ea
where $S^{\a\b}$ is the inverse matrix to $S_{\a\b}$.
Imposing that this vanishes we get the conditions
\ba
S_{MN} & = & M_{M\a} S^{\a\b} M_{\b N} ~ ,
\nonumber\\
S_{ij} & = & S_{i\a} S^{\a\b} S_{\b j} ~ ,
\label{con2}\\
M_{iN} & = & S_{i\a} S^{\a\b} M_{\b N} ~ .
\nonumber
\ea

\no
Having established the on-shell Lorentz invariance of the action it remains to show the Lorentz invariance of the equations of motion  \eqn{eqEm2}. In order to do
that we define a set of projection operators as
\be
(P_\pm)^\m{}_\n = \ha (\d^\m{}_\n \mp S^{\m\l} M_{\l\n})
\label{proojj}
\ee
and their invariant subspaces as
\be
(Q_\pm)^\m{}_i = \ha S^{\m\n} (S_{\n i} \mp M_{\l i}) ~ .
\label{prjj}
\ee
Indeed, using \eqn{con2}, one can readily verify that they satisfy the required properties
\be
P_\pm^2 = P_\pm ~ ,~~~~~ P_\pm P_\mp = 0 ~ , ~~~~~ P_\pm Q_\pm = Q_\pm ~ ,
~~~~~ P_\pm Q_\mp =0 ~ .
\label{proop}
\ee
Then \eqn{eqEm2} with the definition \eqn{eqEm} can be written in the
form
\be
E_\m = S_{\m\n} (E_+^\n + E_-^\n) ~ ,
\label{empm}
\ee
where
\be
E_\pm^\m = (P_\mp)^\m{}_\n \del_\pm X^\n + (Q_\mp)^\m{}_i \del_\pm Y^i ~ .
\label{loreq}
\ee
Then using the properties \eqn{proop} the equations of motion \eqn{eqEm2}
can be easily shown to be equivalent to
\be
E_\pm^\m = 0~ ,
\label{epm}
\ee
which has the required form, since they remain invariant under Lorentz transformations.

\subsubsection{The off-shell symmetry}
In addition to the on-shell Lorentz symmetry \eqn{Lllor} we can construct some
modified Lorentz transformations,  similar to those which appear in a simpler setting in \cite{FloJa}, under which the action is invariant off-shell.
These transform the fields as
\ba
\d X^\m & =  & -\s \del_\tau X^\m - \tau \del_\s X^\m -\s (E^\m_+ + E^\m_-)~ ,
\nonumber \\
\d Y^i & =  & -\s \del_\tau Y^i - \tau \del_\s Y^i ~ .
\label{lornew}
\ea
We see that the $Y^i$'s have the usual global Lorentz transformation rules.
On-shell the same is of course true for the $X^\m$'s as well.

\subsection{Solving the conditions}

\no
Let's introduce a set of vielbeins $e_\m^A$ and their inverses $e^\m_A$. Then we may write
$S_{\m\n}$ in the standard form
\be
S_{\m\n} = e^A_\m e^B_\n \eta_{AB}\ ,
\label{smn}
\ee
where $\eta_{AB}$ is the tangent space metric.
Similarly, let's choose matrices $M_{MN}$ such that
\be
M_{\m\n}  =  - R_{AB} e^A_\m e^B_\n\ ,
\label{smn1}
\ee
for some constant symmetric matrix $R_{AB}$.

\no
Then, the conditions \eqn{con2} are solved by introducing a set of matrices $F^a_i$ that may depend
on the spectator field $Y^i$. We obtain
\ba
S_{\m i} & = & \eta_{AB} e^A_\m F^B_i ~ ,
\nonumber\\
S_{ij} & =  & \eta_{AB} F^A_i F^B_j ~ ,
\nonumber
\\
M_{ij} & = & - R_{AB} F^A_i F^B_j ~ .
\\
M_{\m i} & = & - R_{AB} e^A_\m  F^B_i ~ .
\nonumber
\ea
We may choose for the matrix $C_{ij}$ to equal $S_{ij}$ since its antisymmetric part
is Lorentz invariant by itself.
In addition, $R_{AB}$ obeys the compatibility condition
\be
R_{AC} \eta^{CD} R_{DB} = \eta_{AB} \ .
\label{rac}
\ee
The matrix $R_{AB}$ can be taken to be equal to unity by appropriate choice of the vielbeins.
Then \eqn{rac}
shows that $\eta^{-1}=\eta$, which implies that $\eta$ is diagonal with plus and minus entries \cite{TseDual}.
Having in mind applications to T-duality we take $\eta$ to be traceless as well.

\no
To proceed further, we find it necessary to impose for the matrix $M_{MN}$ the conditions
\ba
\hat \nabla_\m M_{\n\l} & = & 0 ~ ,
\nonumber
\\
\hat \nabla_\m M_{\n i} & = & 0 ~ ,
\label{con41} \\
\del_\m M_{ij} &  = & 0 ~ ,
\nonumber
\ea
for some connection $\Om^\m_{\n\l}$. In the covariant differentiation
the index $i$ is assumed not to transform.
Moreover, we impose that
\be
\hat \G_{\m;\n i} = \hat \G_{j;\n i} =0 \ ,\qq \del_i C_{\m\n} =0\ ,
\label{con5}
\ee
Then the last two terms in \eqn{varact1} are written as
\be
\ha \del_\m M_{N\L}
\del_1 X^N \del_1 X^\L + \hat \G_{N;\m\L} \del_0 X^N \del_1 X^\L =
\left(\Om_{\m\l}^\n M_{\n\L} \del_1 X^N + \hat \G_{N;\m\l} \del_0 X^N\right)\del_1 X^\l\ .
\label{ofh3}
\ee
Identifying the connection as
\be
\Om^\m_{\n\l}= S^{\m\r}\hat \G_{\r;\n\l}\ ,
\ee
we find that the of right hand side of \eqn{ofh3} becomes
\ba
&& \left[\hat \G^\n_{\m\l}(M_{\n N}\del_1 X^N + S_{\n\l} \del_0 X^\l) + \hat \G_{i;\m\l} \del_0 X^i \right]
\del_1 X^\l
\nonumber\\
&& =\hat \G^\n_{\m\l} E_\n \del_1 X^\l
+ (\hat \G_{i;\m\l}-S_{\n i}\hat \G^\n_{\m\l})\del_0 X^i \del_1 X^\l \ .
\label{ahdk}
\ea
From the above and the assumption \eqn{assup}, we see that the form of the connection should be
\be
\hat \G^\n_{\m\l} = \L^\n_A \del_\l \L^A_\m\quad \Longrightarrow\quad
\G^\n_{\l\m} =  \L^\n_A \del_\l \L^A_\m - \ha H^\n{}_{\l\m}\ ,
\ee
in order for the ``small" gauge invariance \eqn{invars} to exist. In the second step
we have written the expression for the usual symmetric Levi--Civita
connection $\G^\n_{\m\l}$ constructed out of
the symmetric tensor $S_{\m\n}$. The above expression implies that
\be
\del_\m \L_\l^A - \del_\l \L_\m^A = \L^A_\n H^\n{}_{\m\l}\ .
\label{rra}
\ee

\no
In addition, the last term in \eqn{ahdk} should vanish, that is
\be
\hat \G_{i;\m\l}= S_{\n i}\hat \G^\n_{\m\l}= S_{\n i} \L^\n_A \del_\l \L^A_\m\ ,
\label{nme3}
\ee
should hold. From \eqn{con5} it is easy to deduce that $C_{\m i}$ is a total derivative.
Hence, with no loss of generality we take
\be
C_{\m i}=0\ .
\ee
Then with the help of \eqn{con5} we write \eqn{nme3} as
\be
\ha \del_\m C_{i\l} = S_{\n i} \L^\n_A \del_\l \L^A_\m\ .
\label{jsdh}
\ee
Finally, note that due to the first of \eqn{con41} and \eqn{smn1} the (torsion free)
spin connection and the torsion are related as\footnote{In our conventions
\be
\hat \nabla_\m V^a_\n = \nabla_\m V^a_\n - \ha H_{\m\n}{}^\l V^a_\l\ ,\qq
\nabla_\m V^a_\n = \del_\m V^a_\n-\Gamma_{\m\n}^\l V^a_\l  + \om_\m{}^a{}_b V^b_\n\
\ee
and $\nabla_\m e^a_\n =0 $.
}
\be
\om_\m{}^{AB} +\ha H_\m{}^{\n\l} e^A_\n e^B_\l = 0 \ .
\label{omh}
\ee

\no
The form of the condition \eqn{rra} suggests that the background corresponds to a group manifold.
With this in mind, we now set up the notation and define the necessary quantities.
Let the $X^\m$'s parametrize an element $G$ of a group $D$. We introduce
a set of representation matrices
$\{T^A\}$, $A=1,2,\dots , \dim(D)$, satisfying
\be
[T^A,T^B]= i f^{AB}{}_C T^C\ .
\ee
Next we introduce the left-invariant Maurer--Cartan forms
\be
L^A_\m = - i {\rm Tr}(T^A G\inv \del_\m G)\ ,
\ee
with the fundamental property, the Maurer--Cartan equation,
\be
\del_\m L^A_\n - \del_\n L^A_\m = f^A{}_{BC} L^B_\m L^C_\n \ .
\label{ff1}
\ee
We also define the right-invariant Maurer--Cartan forms
\be
R^A_\m = - i {\rm Tr}(T^A \del_\m G G\inv)\
\ee
for which
\be
\del_\m R^A_\n - \del_\n R^A_\m =- f^A{}_{BC} R^B_\m R^C_\n \ .
\label{ff2}
\ee
The two forms are related by
\be
R^A_\m = C^A{}_B L^B_\m \ ,
\ee
where the matrix $C^{AB}$ (not related to the one in \eqn{act1}) is defined as
\be
C^{AB}={\rm Tr}(T^AGT^B G^{-1})\ ,
\ee
and indices are raised and lowered with the tangent space metric.
$C^{AB}$ is orthogonal and among the many properties it obeys we will need that
\be
(C^T \del_\m C)^{AB} = f^{AB}{}_C L^C_\m\ .
\label{lablo}
\ee

\no\
Using the above, if we choose as our vielbein the left-invariant Maurer--Cartan forms
\be
e^A_\m = L^A_\m\ ,
\label{ell1}
\ee
we obtain from \eqn{ff1} that the spin connection is
\be
\om_\m^{AB} = \ha f^{AB}{}_C L_\m^C\ .
\label{dhk11}
\ee
Then from \eqn{omh} we have to choose that
\be
H^{AB}{}_C = - f^{AB}{}_C\ ,
\label{dhkk11}
\ee
in tangent space indices. Also from \eqn{rra} and \eqn{ff2},
the matrix $\L^A_\m = R^A_\m$.

\no
Instead, if we choose as our vielbein the right-invariant Maurer--Cartan forms
\be
e^A_\m = R^A_\m\ ,
\ee
then from \eqn{ff2} we get for the spin connection
\be
\om_\m^{AB} =- \ha f^{AB}{}_C R_\m^C\ .
\ee
Then from \eqn{omh} we have to choose that
\be
H^{AB}{}_C = + f^{AB}{}_C\ ,
\ee
in tangent space indices. Also from \eqn{rra} and \eqn{ff1} the matrix $\L^A_\m = L^A_\m$.

\no
Of course quantities and relations
that can be computed using the metric are frame independent. For instance,
the Riemann curvature is
\be
R_{AB;CD} = {1\ov 4} f_{AB}{}^E f_{ECD}\
\ee
and we note the identity
\be
\nabla_\m H_{\n\r\l} = 0 \ ,
\ee
for torsion proportional to the structure constants as in our case. Also in WZW models, the fine tuning
of geometry and torsion gives the identity \cite{Zachos}
\be
\hat R_{\m\n;\r\l}=0\ ,
\ee
where the generalized Riemann curvature tensor that uses $\hat \G^\m_{\n\l}$ as a connection is defined in
\eqn{gee1}.

\no
In what follows we choose to work with the left-invariant forms as our vielbein's, that is with the
choice \eqn{ell1}.

\no
From \eqn{con5} and
\eqn{lablo} we find from \eqn{jsdh} that
\ba
C_{i \m } & = & 2 \eta_{AB} F^A_i e^B_\m  ~  ,
\label{cim}
\ea
where we repeat that the vielbein is given by \eqn{ell1}.

\no
Gathering the previous scattered results we obtain the action with Lagrangian density
\ba
 \cL & = &   \cL_{\rm WZW}  + \ha  \eta_{AB}\left(
F^A_i F^B_j \del_0 Y^i \del_1 Y^j + 2 F^A_i L^B_\m  \del_0 Y^i \del_1 X^\m\right)
\nonumber\\
&&
+ \ha  R_{AB}\left(- L^A_\m L^B_\n \del_1 X^\m \del_1 X^\n
- F^A_i F^B_j \del_1 X^i \del_1 Y^j - 2 L^A_i F^B_i \del_1 X^\m \del_1 Y^i
\right)\ ,
\label{actwzw}
\ea
where $\cL_{\rm WZW}$ corresponds to a WZW model action.
It is not difficult to check that this has
the form of the duality invariant action used for Poisson--Lie T-duality. In particular, one
can verify that by comparing with eq. (B.1) of \cite{PLsfe2}. We emphasize that, unlike the discussions
related to Poisson-Lie T-duality, the group $D$ is not necessarily a Drinfeld double, although some
important applications of our construction will be in that direction, in section 4.

\section{RG flows of Non-Lorentz invariant actions}

In this section we shall explain in detail the computation
of the effective action of the two-dimensional bosonic $\s$-model action \eqn{act1} or equivalently
of \eqn{actwzw}, since we want to have an emergent Lorentz symmetry on-shell. To simplify the
discussion we disregard completely the spectator fields $Y^i$.

\subsection{A brief review of the background field expansion}

In order to compute the quantum anomalies of the actions \eqn{act1} and \eqn{actwzw},
both Weyl and Lorentz ones,
it is necessary to calculate the effective action.
For that purpose we will use the covariant background field method \cite{Alvarez,Zachos}
and in particular the algorithmic method of \cite{Mukhi} and \cite{Howe} to study the
renormalization of this two-dimensional sigma model. Our starting point is to
expand the fields $X^\mu$ around the classical solution $\chi^\mu$ and its fluctuation
$\pi^\mu$ as $X^\mu=\chi^\mu+\pi^\mu$. However, this split leads to a non-covariant
expansion, since $\pi^\mu$ cannot be interpreted as a vector. To achieve covariance,
we use a non-linear split based on the tangent vector $\xi^\mu$ of the geodesic
that connects $\chi^\mu$ and $\chi^\mu+\pi^\mu$. Consider the interval $s\in [0,1]$ and $X^\mu(s)$
such that
\ba
\label{A-2}
X^\mu(0)=\chi^\mu\ ,\qq \del_sX^\mu\big{|}_{s=0}= \xi^\m\ ,\qq X^\mu(1)=\chi^\mu+\pi^\mu\ ,
\ea
which in addition satisfies the geodesic equation
\ba
\label{A-3}
\del^2_sX^\mu+\G_{\nu\k}^\mu\del_sX^\nu\del_sX^\k=0\ ,
\ea
where $\G_{\nu\k}^\mu$ is the standard Levi--Civita connection, built
from a metric. It turns out that the covariant Lagrangian is given
by a Taylor expansion in the parameter $s$, as
\ba
\label{A-4}
{\cal L}(s=1)=\sum_{n=0}^\infty\frac{1}{n!}\del_s^n{\cL}\big{|}_{s=0}=
\sum_{n=0}^\infty\frac{1}{n!}\nabla_s^n{\cL}\big{|}_{s=0}\ ,
\ea
where $\nabla_s$ is the covariant derivative along to the curve $X^\mu(s)$.
For later convenience, we present the formulae
\ba
\label{A-5}
&&\nabla_s\del_s X^\mu=0\ ,\qq
 [\nabla_s,\nabla_a]X^\mu=R_{\k\nu;}{}^\mu{}_\rho X^\rho\del_s X^\k\del_a X^\nu\ ,
\nonumber \\
&&\nabla_s\del_a X^\mu=\nabla_a\del_s X^\mu=\del_a\del_s X^\mu+\G _{\k\l}^\mu\del_a X^\k\del_s X^\l\ ,
\ea
where $\nabla_s$
denotes the worldsheet coordinates from which the fields $X^\mu$ depend, in our case $\a=0,1$.
Now, we are in position to apply this method directly to \eqn{act1} after
we choose what our metric in this expansion would be.
The two possibilities are, the symmetric part of $C_{\mu\nu}$, i.e. $S_{\mu\nu}$ and
$M_{\mu\nu}$.

\no
To calculate the one-loop effective action of \eqn{act1} we have
to perform a path integral over the $\xi$ fluctuations and therefore
we need to calculate the Weyl and Lorentz anomalies of the divergent contributions
of the second order expansion of this action. These come schematically from
the $\xi^2$ and $\xi^4$ contractions. In particular,
these divergences originate from single contractions in $\langle\xi\xi \rangle$ and double
contractions in $\langle \xi\del\xi\xi\del\xi \rangle$. Due to the fact that the candidate metrics
$S_{\mu\nu}$ and $M_{\mu\nu}$ depend on the background fields $\chi^\mu$, it is
convenient to go to the tangent
frame, defined by the vielbeins of our metric.
For our purposes we take
$S_{\mu\nu}$ to be our metric.

\subsection{Background field expansion}

We consider the covariant expansions the two terms in \eqn{act1} separately and
label them as ${\cal S}_{\rm WZW}$ and ${\cal S}_{\rm NL}$, respectively.

\subsubsection{The WZW action}

The WZW part of the Lagrangian \eqn{act1}, can be written as
\ba
\label{A-6}
{\cal S}_{\rm WZW}=\ha\int d\tau d\s\ C_{\mu\nu}\del_0 X^\mu\del_1 X^\nu=
\ha\int d\tau d\s (S_{\mu\nu}+B_{\mu\nu})\del_0 X^\mu\del_1 X^\nu\ ,
\ea
where $S_{\mu\nu}$ and $B_{\mu\nu}$ are the symmetric and antisymmetric parts of
$C_{\mu\nu}$. The first derivative of the action with respect to $s$
after an integration by parts in the term involving the two form potential, has Lagrangian density given by
\be
\label{A-7}
\ha \left(S_{\mu\nu}\nabla_0\del_sX^\mu\del_1X^\nu+\del_0X^\mu\nabla_1\del_sX^\nu
+ H_{\l\m\n} \del_s X^\l\del_0X^\mu\del_1X^\nu\right) \ ,
\ee
where $H_{\k\l\mu}=\del_{[\k}B_{\l\mu]}$ 
and note we used the fact that we have a Levi--Civita connection.
Using that we may compute the second derivative of the action with respect to $s$ and therefore after
setting $s=0$ the second order term in the expansion \eqn{A-4}. We eventually find that
\ba
\label{A-8}
\ha\ {\cal S}_{\rm WZW}^{(2)}\Big{|}_{s=0}&=&\ha\int d\tau d\s\left[
S_{\mu\nu}\nabla_0\xi^\mu\nabla_1\xi^\nu+(R_{\k\mu;\nu\l}+\ha\nabla_\k H_{\l\mu\nu})\xi^\k\xi^\l
\del_0\chi^\mu\del_1\chi^\nu \nonumber \right.\\
&&\left.+\ha H_{\l\mu\nu}\xi^\l(\nabla_0\xi^\mu\del_1\chi^\nu+\del_0\chi^\mu\nabla_1\xi^\nu)\right]\ .
\ea
Equivalently, this can be written in terms of the generalized Riemann tensor that uses $\hat \G^\m_{\n\l}$
as a connection. Recalling that\footnote{In our conventions we antisymmetrize indices as $[ab]=ab-ba$.}
\be
\hat R_{\m\n;\r\l}= R_{\m\n;\r\l} - \ha \nabla_{[\n} H_{\m]\r\l} +{1\ov 4}
H_{\r[\m}{}^\a H_{\n]\l\a}\ ,
\label{gee1}
\ee
we arrive at
\ba
\label{A-9}
\ha\ {\cal S}_{WZW}^{(2)}&=&\ha\int d\tau d\s\left[
S_{\mu\nu}\hat{\nabla}_0\xi^\mu\hat{\nabla}_1\xi^\nu+
\hat{R}_{\k\mu;\nu\l}\xi^\k\xi^\l
\del_0\chi^\mu\del_1\chi^\nu\right]\ ,
\ea
where we have introduced the generalized covariant derivatives
\ba
\label{A-10}
&&\hat{\nabla}_0 \xi^\nu = \nabla_0 \xi^\nu +\ha H_\k{}^\nu{}_\l\ \del_0\chi^\k\xi^\l\ ,
\nonumber \\
&&\hat{\nabla}_1 \xi^\nu = \nabla_1 \xi^\nu -\ha H_\k{}^\nu{}_\l\ \del_1\chi^\k\xi^\l\ ,
\label{hfkk1}
\ea
with
\be
\nabla_a \xi^\mu = \del_a \xi^\m + \G^{\m}_{\n\l} \del_a \chi^\n \xi^\l\ .
\ee
Note that the sign swap in the second terms in \eqn{hfkk1} above indicates that this is not
simply the covariant derivative with generalized connection
pulled back to the world sheet in a naive way.

\no
Using the special form of our background and in particular \eqn{dhk11} and \eqn{dhkk11}
we find from \eqn{A-10} that
\ba
\hat{\nabla}_0 \xi^\rho &=& \nabla_0 \xi^\rho +\ha H_\l{}^\rho{}_\mu\
\del_0\chi^\l\xi^\mu
\nonumber\\
&=&L^\rho_A(\nabla_0\xi^A-\ha f_B{}^A{}_C
L_\nu^B\del_0\chi^\nu\xi^C)
\nonumber\\
&=&L^\rho_A(\del_0\xi^A+\om_\n{}^A{}_C\del_0\chi^\n\xi^C-\ha
f_B{}^A{}_C L_\nu^B\del_0\chi^\nu\xi^C)
\\
&=&L_A^\rho\del_0\xi^A\ ,\label{A-16a}
\nonumber
\ea
where $\xi^\rho = L^\rho_A \xi^A$
and
\ba
\hat{\nabla}_1 \xi^\rho &=& \nabla_1 \xi^\rho -\ha H_\l{}^\rho{}_\mu\
\del_1\chi^\l\xi^\mu
\nonumber\\
&=&L^\rho_A(\nabla_1\xi^A+\ha f_B{}^A{}_C
L_\nu^B\del_1\chi^\nu\xi^C)
\nonumber\\
&=&L^\rho_A(\del_1\xi^A+\om_\n{}^A{}_C\del_1\chi^\n\xi^C+\ha
f_B{}^A{}_C L_\nu^B\del_1\chi^\nu\xi^C)
\\
&=&L_A^\rho\left(\del_1\xi^A+f^A{}_{BC}\xi^B L^C_\l\del_1\chi^\l\right)\ . \label{A-16b}
\nonumber
\ea
Note that these expressions depend crucially on the relative sign between the
torsion and the spin connection. In the analysis of a standard WZW either
sign choice is consistent and leads to the same ultimate result.
However, in our case of non-Lorentz invariant theories we found that
we had to specify opposite signs in \eqn{omh} to ensure on shell invariance.

\no
Finally, from Eq.\eqn{A-9}, \eqn{A-16a} and \eqn{A-16b} we obtain that
\ba
\label{A-17}
\ha\ {\cal S}_{\rm WZW}^{(2)}=\ha\int d\tau d\s \left[
\eta_{AB}\del_0\xi^A\del_1\xi^B-f_{ABC}\xi^A\del_0\xi^B L^C_k\del_1\chi^k\right] .
\ea
Clearly, upon exponentiation and wick contraction there is no divergence.
This can be easily seen from \eqn{A-17} since, when we only have the WZW model kinetic term the propagator is
irregular
in the 01 directions and proportional to $\eta^{AB}$.
Thus, the $\xi\del_0\xi$ term vanishes, whereas the $(\xi\del_0\xi)^2$ is
finite.
This is the familiar statement that the WZW model is not renormalized and is conformal.
However, as we will now come onto, when we also consider the interaction term ${\cal S}_{\rm NL}$
we modify the form of the propagators and the exponentiation of \eqn{A-17}
does provide an important contribution to both the Weyl divergence and the Lorentz anomaly.

\subsubsection{The NL action}

The second term of the Lagrangian \eqn{act1}, is
\ba
\label{A-18}
{\cal S}_{\rm NL}=\ha\int d\tau d\s M_{\mu\nu}\del_1 X^\mu\del_1 X^\nu\ .
\ea
The first derivative of the action with respect to $s$ has Lagrangian density given by
\ba
\label{A-19}
\ha \left(\nabla_\k M_{\mu\nu}\del_s X^\k\del_1X^\mu X^\nu+
2M_{\mu\nu}\nabla_1\del_s X^\mu\del_1X^\nu\right)\ .
\ea
Similarly, to \eqn{A-8} we find that the second order term in the expansion \eqn{A-4} is
\ba
\label{A-20}
\ha\ {\cal S}_{\rm NL}^{(2)}\big{|}_{s=0}&=&\ha\int d\tau d\s\left[
M_{\mu\nu}\nabla_1\xi^\mu\nabla_1\xi^\nu+(R_{\k\mu;\nu\l}+\ha\nabla_\k \nabla_\l M_{\mu\nu})\xi^\k\xi^\l
\del_1\chi^\mu\del_1\chi^\nu\right. \nonumber\\
&&\left. + 2 \nabla_k M_{\mu\nu}\xi^\k\nabla_1\xi^\mu\del_1\chi^\nu\right]\ .
\ea
Expanding this using $M_{\mu\nu}=-L^A_\mu L^B_\nu R_{AB}$ and a lengthly procedure we obtain an analogous
to \eqn{A-17} action for the fluctuations.

\subsection{Calculation of the effective action}

The action for the fluctuations of the complete action \eqn{act1}
can be written as a sum of a kinetic and interacting terms for the $\xi^A$'s, as
\be
\label{A-21a}
S^{(2)}=S_{\rm kin}+S_{\rm int}\ ,
\ee
where
\be
S_{\rm kin}=\ha\int d\s d\tau\left(
\eta_{AB}\del_0\xi^A\del_1\xi^B-R_{AB}\del_1\xi^A\del_1\xi^B\right)\ ,
\ee
and
\be
S_{\rm int}=\ha\int d\s d\tau\left(
I_{AB}\xi^A\xi^B+J_{AB}\xi^A\del_1\xi^B+K_{AB}\xi^A\del_0\xi^B\right) \ ,
\ee
with
\ba
&&I_{AB}=-\ha\del_1\chi^\mu\del_1\chi^\nu L^C_\mu L^D_\nu\left[
f_{AC}{}^E f_{BD}{}^FR_{EF}+
(2f_{AF}{}^ER_{EC}+f_{AC}{}^ER_{EF})f_{BD}{}^F\right]\ ,
\nonumber \\
&&J_{AB}=(f_{BA}{}^CR_{CE}+2f_{EA}{}^CR_{CB})L_\mu^E\del_1\chi^\mu\ ,
\label{A-22}\\
&&K_{AB}=-f_{ABC}L_\mu^C\del_1\chi^\mu\ .
\nonumber
\ea
We note that the terms labeled by $I_{AB}$ and $J_{AB}$ come from the ${\cal S}^{(2)}_{\rm NL}$ action,
while the term $K_{AB}$
comes from the ${\cal S}^{(2)}_{\rm WZW}$ action.

\no
The effective action $S_{\rm eff}(\chi)$ at one loop, it is given in terms of
the interacting Lagrangian as
\ba
\label{A-23}
e^{iS_{\rm eff}(\chi)}=\int[D\xi]e^{iS^{(2)}[\chi,\xi]}\quad \Longrightarrow \quad
S_{\rm eff}(\chi)=\langle S_{\rm int}\rangle +
\frac{i}{2} \langle S_{\rm int}^2 \rangle\ .
\ea
Substituting \eqn{A-21a} in \eqn{A-23} we find that the effective Lagrangian is given by
\ba
\label{A-24}
S_{\rm eff}(\chi)&=&\ha\int d\s d\tau(q_1+q_2+q_3+q_4)
\nonumber\\
&=&\ha\int d\s d\tau\left(I_{AB} \langle\xi^A\xi^B \rangle
+ \frac{i}{4}J_{AB}J_{CD} \langle \xi^A\del_1\xi^B\xi^C\del_1\xi^D \rangle
\right.
\nonumber\\
&& +\frac{i}{4}K_{AB}K_{CD} \langle \xi^A\del_0\xi^B\xi^C\del_0\xi^D \rangle +
\\
&&+ \left.\frac{i}{4}(J_{AB}K_{CD}+K_{AB}J_{CD}) \langle \xi^A\del_1\xi^B\xi^C\del_0\xi^D \rangle \right)\ ,
\nonumber
\ea
where $q_i, i=1,2,3,4$ are respectively the first, second, third and fourth term in the above expression.
As for the $\xi^2$ contractions and $(\xi\del\xi)^2$, these can be found from the
$S_{\rm kin}$ and were computed in detail in the appendix of \cite{Berman1} and will not be repeated here.
The end results are
\ba
\langle \xi^A\xi^B \rangle & = & R^{AB}\D(0)+\eta^{AB}\Th(0)\ ,
\nonumber\\
i\langle \xi^A\del_1\xi^B\xi^C\del_1\xi^D \rangle & \simeq & \ha(R^{A[C}R^{D]B}-\eta^{A[C}\eta^{D]B})\D(0)\ ,
\nonumber\\
i\langle \xi^A\del_0\xi^B\xi^C\del_0\xi^D \rangle & \simeq & -\ha(R^{A[C}R^{D]B}+3\eta^{A[C}\eta^{D]B})\D(0)
\label{A-25}\\
&& \phantom{}
-(R^{A[C}\eta^{D]B}+\eta^{A[C}R^{D]B})\Th(0)\ ,
\nonumber\\
\langle \xi^A\del_0\xi^B\xi^C\del_1\xi^D \rangle &  \simeq & -\ha(R^{A[C}\eta^{D]B}+
\eta^{A[C}R^{D]B})\D(0)-\eta^{A[C}\eta^{D]B}\Th(0)\ ,
\nonumber
\ea
where $\D(0)$ and $\Th(0)$ are the Weyl and Lorentz anomalies, respectively and $\simeq$
denotes the fact that we have kept only the terms relevant to the Lorentz and the
 Weyl divergence.\footnote{In these expressions $\langle \xi^A\xi^B \rangle$
is really short hand for  $\int d^2 \sigma' \langle \xi^A(\sigma) \xi^B(\sigma') \rangle $.
}

\subsubsection{Weyl anomaly}

We shall now compute the Weyl anomaly of the effective action
by plugging \eqn{A-22} and \eqn{A-25} in \eqn{A-24}. The result can be written as
\ba
q_1^W&=&-\ha\del_1\chi^\mu\del_1\chi^\nu L^C_\mu L^D_\nu\left(
f_{AC}{}^E f_{BD}{}^F R_{EF}R^{AB}+\right.
\nonumber \\
&&\left.(\ 2f_{AF}{}^E R_{EC}+f_{AC}{}^E R_{EF})f_{BD}{}^F R^{AB}\right)\ ,\nonumber \\
q_2^W&=&\frac{1}{4} f_{AB}{}^C f_{DE}{}^F R_{CE}R_{FK}(R^{AD}R^{BE}-\eta^{AD}\eta^{BE})
\del_1\chi^\mu\del_1\chi^\nu L^E_\mu L^K_\nu\ ,
\label{A-26}\\
q_3^W&=&-\frac{1}{4}\left(f_{ABC}f_{DEF}R^{AD}R^{BE}+3f_{ABC}f^{AB}{}_F\right)
\del_1\chi^\mu\del_1\chi^\nu L^C_\mu L^F_\nu\ ,\nonumber \\
q_4^W&=&\left(f_{ABC}f^{AB}{}_D+f_{CE}{}^F R_{FB} f_{DA}{}^B R^{EA}+f_{ABC}f^{AF}{}_E R_{FD}R^{EB}\right)
\del_1\chi^\mu\del_1\chi^\nu L^C_\mu L^D_\nu\ ,\nonumber
\ea
where we
note that we have made heavy use of the compatibility condition
\eqn{rac} in the calculation of $q_2$.
Adding up the contributions we find that the Weyl anomaly reads
\ba
\label{A-27}
{\rm Weyl}&=&\frac{1}{4} f_{AB}{}^Cf_{DE}{}^F\left(R_{CK}R_{FH}R^{AD}R^{BE}-
R_{CK}R_{FH}\eta^{AD}\eta^{BE}\right.
+\nonumber\\
&&\left.\eta_{CK}\eta_{FH}\eta^{AD}\eta^{BE}-\eta_{CK}\eta_{FH}R^{AD}R^{BE}\right)
L^K_\mu L^H_\nu\del_1\chi^\mu\del_1\chi^\nu
\nonumber\\
&=&\frac{1}{4}
(R_{AC}R_{BF}-\eta_{AC}\eta_{BF})
(R^{KD}R^{HE}-\eta^{KD}\eta^{HE})
f_{KH}{}^Cf_{DE}{}^F \times
\\
&&L^A_\mu L^B_\nu\del_1\chi^\mu\del_1\chi^\nu\ .
\nonumber
\ea
However, the zeroth order in the expansion of ${\cal S}_{\rm NL}$ has the Lagrangian density
\ba
\label{A-28}
M_{\mu\nu}\del_1\chi^\mu\del_1\chi^\nu=-
R_{AB}L^A_\mu L^B_\nu\del_1\chi^\mu\del_1\chi^\nu\ .
\ea
Thus, since the Weyl anomaly counter terms can be absorbed as a redefinition of the $R_{AB}$
we find that the action \eqn{act1} is renormalizable at one loop.
Moreover the RG flow of  $R_{AB}$ can be directly read off from \eqn{A-27}
and it is equal to
\ba
\label{A-29}\boxed{
\frac{dR_{AB}}{dt}=\frac{1}{4}
(R_{AC}R_{BF}-\eta_{AC}\eta_{BF})
(R^{KD}R^{HE}-\eta^{KD}\eta^{HE})
f_{KH}{}^Cf_{DE}{}^F}\ ,
\ea
where $t=\ln \m$, with $\m$ being the energy scale.
This is a quite simple formula and constitutes one of the main
results of present paper. One can
readily check that it is compatible with the condition \eqn{rac}.\footnote{It is
interesting and quite curious that the right hand side
of \eqn{A-29} appeared before in a study of
generalizations of WZW-type conformal models \cite{tse93}.}

\subsubsection{Lorentz anomaly}

We shall now compute the Lorentz anomaly of the effective action
by plugging \eqn{A-22} and \eqn{A-25} in \eqn{A-24}. The result can be written as
\ba
q_1^L&=&-\ha\del_1\chi^\mu\del_1\chi^\nu L^C_\mu L^D_\nu\left(
f_{AC}{}^E f_{BD}{}^F R_{EF}\eta^{AB}+\right.
\nonumber
\\
&&\left.(\ 2f_{AF}{}^E R_{EC}+f_{AC}{}^E R_{EF})f_{BD}{}^F \eta^{AB}\right)\ ,\nonumber \\
q_2^L&=&0\ ,
\label{A-30}
\\
q_3^L&=&f_{ABC}f_{DE}{}^B R^{AD}L^C_\mu L^E_\nu\del_1\chi^\mu\del_1\chi^\nu\ ,\nonumber\\
q_4^L&=&(-f_{AB}{}^C R_{CE}+2f_{EA}{}^C R_{CB})f^{AB}{}_D L^E_\mu L^D_\nu\del_1\chi^\mu\del_1\chi^\nu\ .\nonumber
\ea
Adding up the contributions we find that the Lorentz anomaly is zero. Thus, the system
is Lorentz invariant at one loop and this represents an important and non-trivial demonstration of consistency.

\section{Application to Poisson--Lie T-duality }

Let's concentrate to the case where the group $D$ providing the symmetry structure in our
construction is the Drinfeld double of
two groups ${\cal G}$ and $\tilde{ {\cal G}}$ of equal dimension $d_g$.
Splitting the index $A=(a,\tilde a)$ and using the convention $T^{\tilde a}= \tilde T_a$,
the commutation relations can be written as
\ba
&& [T_a,T_b]  =  i f_{ab}{}^c T_c \ ,
\nonumber\\
&& [\tilde T^a, \tilde T^b]  =  i \tilde f^{ab}{}_c \tilde T^c \ ,
\label{jh9}
\\
&& [T_a,\tilde T^b ] = i \tilde f^{bc}{}_a T_c - i f_{ac}{}^b \tilde T^c\ .
\nonumber
\ea
It turns out that starting from the action \eqn{actwzw} (with no spectators)
and parameterizing the group element $G\in D$ in
two inequivalent ways, namely as $G=\tilde h g $ and $G=h \tilde g$, where $h,g\in {\cal G}$ and
$\tilde h,\tilde g\in\tilde{ {\cal  G}}$, we may solve some constraint type equations \cite{DriDou}
for the fields in $\tilde h$ or those in $h$ and obtain Lorentz-invariant $\s$-model actions of the form
\be
S={1\ov 2} \int Q_{\m\n} \del_+ X^\m \del_- X^\n\ ,
\label{fjll}
\ee
for the remaining fields in $g$ or in $\tilde g$,
respectively.

\no
In the absence of spectator fields, these dual two-dimensional $\s$-model actions
are \cite{KliSevI}
\be
S= {1\ov 2 } \int E_{ab} L^a_\m L^b_\n \del_+ X^\m \del_- X^\n\ ,\qq E= (M - \Pi)^{-1}\ ,
\label{action1}
\ee
and
\be
\tilde S= {1\ov 2 } \int \tilde E^{ab} \tL_{a\m} \tL_{\b\n} \del_+
 \tilde X^\m \del_- \tilde X^\n\ ,\qq \tilde E=(M^{-1} - \tilde \Pi)^{-1}\ ,
\label{action2}
\ee
where $M$ is an arbitrary square matrix of $\dim({\cal G})$
and $\Pi_{ab}$ ($\tilde \Pi^{ab}$)
is an $X^\m$ ($\tilde X^\m$)-dependent
matrix which subtly encodes the group structure of the Double, but whose details
are not relevant in our considerations.

\no
The one-loop beta-functions are expressed as
\ba
{da^i\ov dt}= { 1 \ov \pi} a^i_1\ ,
\ea
where the $a^i_1$'s are chosen such that
\be
\ha \hat R_{\m\n} = \del_{a^i} Q^+_{\m\n} a^i_1 \ ,
\label{1loop}
\ee
where in our case the couplings are the entries of the matrix $M$ in \eqn{action1} and \eqn{action2}.
Also, $\hat R_{\m\n}$ are the components of
the generalized Ricci tensor defined with the connection $\hat \G^{\m}_{\n\r}$ that includes the torsion.
The counter-terms were computed in \cite{Zachos, vanderven,Osborn} and we have also omitted possible field
renormalizations as they are not needed in our case.

\no
To present the result for the beta function equations
we recall the notation introduced in \cite{KSsquarePL}.
We first define the matrices
\be
A^{ab}{}_{c} = \tilde f^{ab}{}_c - f_{cd}{}^a M^{db}\ ,\qq
B^{ab}{}_{c} = \tilde f^{ab}{}_c + M^{ad}f_{dc}{}^b \ ,
\label{fhh11}
\ee
as well as their duals
\be
\tilde A_{ab}{}^{c} =  f_{ab}{}^c - \tilde f^{cd}{}_a (M^{-1})_{db}\ ,\qq
\tilde B_{ab}{}^{c} = f_{ab}{}^c + (M^{-1})_{ad}\tilde f^{dc}{}_b \ ,
\ee
Using these we construct also
\ba
L^{ab}{}_c & = &  \ha (M_s^{-1})_{cd}\left( B^{ab}{}_e M^{ed} + A^{db}{}_e M^{ae}- A^{ad}{}_eM^{eb}
  \right) \ ,
\nonumber\\
R^{ab}{}_c & = & \ha (M_s^{-1})_{cd}\left( A^{ab}{}_e M^{de} + B^{ad}{}_eM^{eb} - B^{db}{}_e M^{ae}
\right) \
\label{kkll1a}
\ea
and
\ba
\tilde L_{ab}{}^c & = &  \ha (\tilde M_s^{-1})^{cd}\left(
\tilde B_{ab}{}^e (M^{-1})_{ed} + \tilde A_{db}{}^e (M^{-1})_{ae}- \tilde A_{ad}{}^e (M^{-1})_{eb}
  \right) \ ,
\nonumber\\
\tilde R_{ab}{}^c & = & \ha (\tilde M_s^{-1})^{cd}\left( \tilde A_{ab}{}^e (M^{-1})_{de}
+ \tilde B_{ad}{}^e (M^{-1})_{eb} - \tilde B_{db}{}^e (M^{-1})_{ae}
\right) \ ,
\label{kkll1}
\ea
where
\be
M_s = \ha (M+M^T) \ ,\qq \tilde M_s = \ha \left[(M^{-1}) + (M^{-1})^T\right]\ .
\ee

\no
The one-loop RG flow system of equations corresponding to \eqn{action1} can be read from
\cite{ValCli}. In the notation of \cite{KSsquarePL} it reads
\be
{d M^{ab}\ov dt}
= R^{ac}{}_d L^{db}{}_c\ .
\label{rg1}
\ee
Similarly, for its dual \eqn{action2} we have
\be
{d (M^{-1})_{ab}\ov dt}
= \tilde R_{ac}{}^d \tilde L_{db}{}^c\ ,
\label{rg2}
\ee
up to a constant overall factor absorbed into a redefinition of $t$.

\no
Developing certain identities among the various quantities defined above it has been shown
in \cite{KSsquarePL} that the two systems \eqn{rg1} and \eqn{rg2} are in fact equivalent.
Therefore, at one-loop in perturbation theory, general $\s$-models related
by Poisson--Lie T-duality are equivalent under the renormalization group flow a fact that seems to
be related to the fact that the $\s$-model actions \eqn{action1} and \eqn{action2} are on
fact canonically equivalent in phase space \cite{PLsfe1,PLsfe2}.

\no
The equivalence of the systems \eqn{rg1} and \eqn{rg2}
does not necessarily mean that the beta functions one would have computed for the
original action \eqn{act1} using \eqn{A-29} would have been the same. The reason is that
as we have mentioned above certain constraints were solved in order to obtain \eqn{action1} and
\eqn{action2} and quantization before
and after solving them might not be commuting operations. In order to check this issue we need to
conveniently parametrize the matrix $R_{AB}$ introduced in \eqn{smn1} in the case that the group
is a Drienfeld double.
First we define an inner product as
\be
\langle T_A|T_B\rangle = \eta_{AB} =  \left(\begin{array}{cc}
  \bf{0} & \mathbb{I}_{d_g\times d_{g}} \\
  \mathbb{I}_{d_g\times d_g} & \bf{0} \\
\end{array}\right)
\label{fh23}
\ee
and then the basis
\be
R^+_a = T_a + M^{-1}_{ab} \tilde T^b\ ,\qq R^-_a = T_a - M^{-1}_{ba} \tilde T^b\ .
\ee
We easily verify that
\be
\langle R^\pm_a|R^\pm_b \rangle = \pm 2 \tilde M_s \ ,\qq \langle R^\pm_a|R^\mp_b \rangle =0 \ ,
\ee
as well as completeness
\be
\ha |R^+_a\rangle (\tilde M_s^{-1})^{ab} \langle R^+_b| -\ha |R^-_a\rangle (\tilde M_s^{-1})^{ab} \langle R^-_b|
=\mathbb{I}\ .
\ee
Then we define the operator
\be
R = \ha |R^+_a\rangle (\tilde M_s^{-1})^{ab} \langle R^+_b| + \ha |R^-_a\rangle (\tilde M_s^{-1})^{ab} \langle R^-_b|\ ,
\ee
whose inner products
\be
R_{AB}= \langle T_A |R|T_B\rangle \ ,
\ee
provide the necessary matrix elements for the computation in \eqn{A-29}. Explicitly, in a block
diagonal form, they are given by
\be
R_{AB} = \left(
           \begin{array}{cc}
             \tilde M_s - B \tilde M_s^{-1}B & -B \tilde M_s^{-1}  \\
             \tilde M_s^{-1} B & \tilde M_s^{-1} \\
           \end{array}
         \right)\ ,
\label{rab}
\ee
where $B= \ha \left[M^{-1}-(M^{-1})^T\right]$.

\no
Before we present some examples we cannot refrain from noting the complexity
of the systems \eqn{rg1} and \eqn{rg2} as compared to
the manifestly duality invariant system \eqn{A-29}.

\no

\subsection{Examples}

We have not attempted to prove in general that in the case of Drinfeld doubles the
RG flow system of equations \eqn{A-29} reduce to \eqn{rg1} (equivalently \eqn{rg2}).
Instead, we will present three non-trivial examples in which we will explicitly compute the beta functions equations
using \eqn{A-29} and recover the same equations that were previously computed using \eqn{rg1}.

\subsubsection{ Semi-Abelian Doubles }
As a first example we consider the case when the commutator relations for the double are given as
\ba
&& [T_a,T_b]  =  i f_{ab}{}^c T_c \ ,
\nonumber\\
&& [\tilde T^a, \tilde T^b]  =  0 \ ,
\\
&& [T_a,\tilde T^b ] = - i f_{ac}{}^b \tilde T^c\ .
\nonumber
\ea
This is know as the semi-abelian double since $\tilde{ {\cal G}} = U(1)^d$
and coincides with the regular notion non-abelian T-duality.
We leave the group ${\cal G}$ general but to keep the
problem simple we consider a point in the coupling space where
\be
M^{ab}  = \kappa \delta^{ab}\ .
\ee
Then from \eqn{kkll1a} we find that $\displaystyle R^{ab}{}_c = - L^{ab}{}_c= {\kappa\ov 2} f^{ab}{}_c $. Then,
from the general RG equations \eqn{rg1} we find the classical result
\be
\frac{ d \kappa} {dt} = - \frac{C_2}{4}\ \kappa^2  \ ,
\ee
where $C_2$ is the quadratic Casimir in the adjoint representation.
In the duality invariant expressions we have that
$R_{AB} = {\rm diag} ( \frac{1}{\kappa} \delta_{ab}, \kappa \delta^{ab} )$
 and can simply calculate the RG equation for $\kappa$ using \eqn{A-29}.
 We find that the above result is recovered.
\subsubsection{ A six-dimensional Drinfeld double}

In this example we use a six-dimensional Drinfeld double based on the
three-dimensional Lie algebras, IX for $\cal{G}$ and V for $\tilde{\cal{G}}$ in the Bianchi classification.
The corresponding generators are
$T_a$ and $\tilde{T}^a$, where $a=1,2,3$. It is convenient to
split the index
$a=(3,\a)$, with $\a=1,2$. The non-vanishing commutation relations are
\ba
&& [T_a,T_b]= i \e_{abc} T_c \ ,\qq [\tilde T_3,\tilde T_\a]= i \tilde T_\a \ ,
\nonumber\\
&&[T_\a,\tilde{T}_\b]=i \e_{\a \b}\tilde{T}_3-i \d_{\a \b}T_3\ ,
\quad [T_3,\tilde{T}_\a]=i\e_{\a \b}\tilde{T}_\b\ ,
\quad [\tilde{T}_3,T_\a]=i \e_{\a \b}\tilde{T}_\b-i T_{\a} \ ,
\label{commu1}
\ea
where $\d_{\a\b}, \e_{\a\b}$ are the Kronecker delta and the antisymmetric
symbol in two dimensions and
$\e_{abc}$ is the totally antisymmetric tensor in three dimensions,
$\e_{123}=1$. From these one can read off the structure constants.


\no
As for the
matrix $R_{AB}$,
as it was previously shown it can be constructed in terms the symmetric and
antisymmetric part of $M^{-1}$,
where $M$ is taken for simplicity to be
\be
\label{Rmatrixb}
M=\left( \begin{array}{ccc}
     a & 1-b & 0 \\
     b-1 & a & 0 \\
     0 & 0 & \frac{a}{1+g} \\
   \end{array}
\right) \, .
\ee
Then from \eqn{rab}
\be
R=\left(\begin{array}{cccccc}
    \frac{1}{a} & 0 & 0 & 0 & \frac{1-b}{a} & 0 \\
    0 & \frac{1}{a} & 0 & \frac{b-1}{a} & 0 & 0 \\
    0 & 0 & \frac{1+g}{a}& 0 & 0 & 0 \\
    0 & \frac{b-1}{a} & 0 & \frac{a^2+(b-1)^2}{a} & 0 & 0 \\
    \frac{1-b}{a} & 0 & 0 & 0 & \frac{a^2+(b-1)^2}{a} & 0 \\
    0 & 0 & 0 & 0 & 0 & \frac{a}{1+g} \\
  \end{array}\right)\ .
\label{Rmatrix}
\ee
Whereas, the matrix $\eta_{AB}$ is given by \eqn{fh23} with $d_g=3$.
Plugging the structure constants and  \eqn{Rmatrix} into \eqn{A-29}
we find
\ba
\label{6dim}
&&{da\ov dt}={1+a^2-b^2\ov 2a^2}((g-1)a^2+(g+1)(b^2-1))\ ,
\nonumber \\
&&{db\ov dt}={b\ov a}(a^2(g-1)+(g+1)(b^2-1))\ ,\\
&&{dg\ov dt}={1+g \ov a}(g(1+a^2)+(g+2)b^2)\ ,
\nonumber
\ea
which is precisely the same system one derives using \eqn{rg1} or \eqn{rg2} \cite{PLsfe4, KSsquarePL}.

\subsubsection{ A sixteen-dimensional Drinfeld double}

In this final example we study a sixteen-dimensional Drinfeld double group
based on an $SU(3)$ group with generators $T_a$ and an abelian
eight-dimensional group
with generators $\tilde{T}^a$, where $a=1,2,\dots,8$.
For the $SU(3)$ group we use the structure constants in the Gell-Mann
basis (see for instance, eq.(5.2) of \cite{MuellerHoissen})
\ba
\label{strSU(3)}
f^{12}{}_3=2,\quad
f^{14}{}_7=-f^{15}{}_6=f^{24}{}_6=f^{25}{}_7=f^{34}{}_5=-f^{36}{}_7=1,\quad
f^{45}{}_8=f^{67}{}_8=\sqrt{3}\ ,
\ea
where the rest are obtained by antisymmetrization in the three indices.
Note that there in an $SU(2)$ subgroup generated by $T^i,\ i=1,2,3$.
The non-vanishing structure constants are read by comparing the first and third equation of \eqn{jh9}.
In particular, we find
that
$f^{a\tilde{b}}{}_{\tilde{c}}=f^{ab}{}_c$ and antisymmetry in the first
two indices is understood. As for the matrix $R_{AB}$, it is built
in terms of the symmetric and antisymmetric part of $M^{-1}$,
where $M$ is taken for simplicity to be a diagonal eight-dimensional matrix
\ba
M=\left(\frac{1}{g}\mathbb{I}_{3\times 3}\right)\oplus
\left(a\ \mathbb{I}_{4\times 4}\right)\oplus b\ .
\ea
The result is a diagonal sixteen-dimensional matrix given by
\ba
\label{Rmatrix2}
R=
\left(g\mathbb{I}_{3\times 3}\right)\oplus
\left(\frac{1}{a}\mathbb{I}_{4\times 4}\right)\oplus\frac{1}{b}\oplus
\left(\frac{1}{g}\mathbb{I}_{3\times 3}\right)\oplus
\left(a\mathbb{I}_{4\times 4}\right)\oplus b\ .
\ea
Whereas the matrix $\eta_{AB}$ is given by \eqn{fh23} with $d_G=8$.
Plugging the structure constants, \eqn{fh23}, \eqn{Rmatrix2} into
\eqn{A-29} we find
\ba
\frac{da}{dt}=\frac{3a^2}{2}\frac{a(bg+1)-4b}{b}\ ,\quad
\frac{db}{dt}=-3a^2\ ,\quad
\frac{dg}{dt}=g^2a^2+2\ ,
\ea
which were precisely the expressions in eq.(5.9) of \cite{KSsquarePL}.

\section{Conclusions}
We showed that demanding on-shell Lorentz invariance highly
constrains the structure of a general Lorentz non-invariant action.
The resulting theories have an underlying group structure and consist
of a WZW term together with some interaction term.   In the case that
the group is a Drinfeld Double these theories are \PL\ invariant $\s$-models.
 By using a background field method we calculated the one-loop effective
 action of these models and found that they were renormalizable and that
a possible quantum Lorentz anomaly vanished.  This is an important consistency
condition of such models.   We also obtained the renormalization group
equations for the couplings of the interaction term in our model.
For the \PL\ invariant theories this provides a duality invariant
description of the RG equations, a key motivator for this work.
 We also verified that for specific examples of the Drinfeld Double
that these duality invariant RG equations agreed with those obtained
from either of the T-dual standard $\s$-models. A very interesting open problem
is to prove this agreement in full generality, although we believe that the non-trivial
examples that we have given leave not much doubt that this will be the case.

\no
An observation, that hints at the utility of
a duality invariant framework, is that duality invariant
RG equations can be computed using simple contractions of structure constants
with constant matrices.
When dealing with the standard $\s$-model one obtains these equations
only by calculating the generalized curvature of what are, in general, extremely complicated
target space backgrounds.

\vskip  .2 cm

\no{\bf Note added:}
Towards the completion of typing the present paper, the work of \cite{avramis} appeared
where similar issues are discussed,
albeit from a different point of view and motivation. The result \eqn{A-29} has also been
derived in that work as well. We acknowledge that
the authors communicated to us their results a few days before submission.

\vskip  .3 cm

\centerline{\bf Acknowledgments}

\no
We would like to thank S. Avramis, D. Berman, R. Reid-Edwards, N. Prezas and G. Travaglini
for helpful discussions. K. Siampos, also thanks G. Papadopoulos for discussions on the background field
method in the Spring of 2008 during his stay at the department of Mathematics at Kings
College London supported by the ``European Superstring Network" with contract MCFH-2004-512194.
In addition, K. Siampos acknowledges support by the  Greek State Scholarship Foundation (IKY) and
D.C.T. is supported by a STFC studentship.


\end{document}